\documentclass[12pt]{article}
\usepackage[utf8]{inputenc}
\usepackage[english]{babel}
\usepackage{amssymb,graphicx}
\usepackage{amsmath,amsfonts}
\usepackage{fullpage}
\def\e{{\rm e}}
\def\d{\partial}
\def\l{\left(}
\def\r{\right)}
\newcommand{\be}{\begin{equation}}
\newcommand{\ee}{\end{equation}}

\newcommand{\ch}{\mathop{\rm ch}\nolimits}
\newcommand{\sh}{\mathop{\rm sh}\nolimits}

\author{Petr Satunin\thanks{{\bf e-mail}: satunin@ms2.inr.ac.ru}
\vspace{.2cm}\\
\normalsize\it Institute for Nuclear Research of the Russian Academy of Sciences, \\ 
      \normalsize \it  60th October Anniversary Prospect, 7a, 117312  Moscow, Russia} 
\title{
\vspace{-2cm}
\begin{flushright}  
\footnotesize{INR-TH/2014-019}
\end{flushright}
\vspace{1cm}
A study of neutral particle decay in magnetic field with the ``Worldline Instanton'' approach}
\date{}
\begin{document}
\maketitle

\begin{abstract}
We study the process of neutrino decay to electron and $W$-boson in the external magnetic field using the semiclassical ``worldline instanton'' approach. Being interested only in the leading exponential factor, we make calculations in a toy model, treating all particles as scalars. This calculation determines the effective threshold energy of the reaction as a function of the magnetic field. Possible astrophysical applications are discussed. It is emphasized that the method is general and is applicable to a decay of an arbitrary neutral  particle into charged ones in the external electromagnetic field.
\end{abstract}

\section{Introduction}

Processes of neutral particle decays in the external magnetic field may be relevant for astrophysics. An important example is the process of photon decay to electron-positron pair in the geomagnetic field, which must be taken into account in the search for cosmic ray photons of ultra high energies \cite{Risse:2007sd}. Other examples include decays of high energetic neutrinos in strong magnetic fields.

Recent detection of very high energy (up to $10^{15}$ eV) neutrinos in the IceCube experiment \cite{Aartsen:2013bka} can open a new branch of astrophysics --- very high energy neutrino astronomy. Although the angular resolution of neutrino detectors is rather low, it will significantly improve in the near future, and may reach the level sufficient for identification of neutrino sources. Therefore, a theoretical study of astrophysically relevant neutrino processes become well motivated.

One class of potential neutrino sources are pulsars and magnetars. These objects generically have a superstrong magnetic field around them in a radius of several kilometres. This raise a question: Can neutrinos escape the region of strong magnetic field if they are produced inside it?  Neutrino dispersion relation in the external magnetic field is modified \cite{Erdas:1990gy}, so its decay, forbidden in the absence of the field, becomes allowed. The main decay channels are $\nu\to\nu e^+e^-$ and $\nu\to e^-W^+$. 

These processes have been studied in many works \cite{Borisov:1985ha,Erdas:2002wk,Bhattacharya:2008px,Kuznetsov:2010sn,Borisov:1993rba,Kuznetsov:1996vy,Kuznetsov:2000gn}  (see also the book \cite{Kuznetsov:2013sea}). The dependence of their widths on the magnetic field and neutrino energy exhibits the following common feature. At small fields or energies they are exponentially suppressed while when the energy or magnetic field exceed certain values (different for different processes) the suppression disappears. In other words, the above reactions proceeds effectively only above a certain threshold energy, which depends on the value of the magnetic field.

The reaction
\be\label{reaction}
\nu \to e^-W^+,
\ee
being of the first order in the weak coupling constant, gives the leading contribution to the neutrino decay width once the energy exceeds the corresponding threshold.

It was analysed for subcritical magnetic fields\footnote{The critical, or Schwinger, magnetic field is obtained as $H_{cr} \equiv m_e^2/e \simeq 4\cdot 10^{13}\;\mbox{G}$, where $m_e$ and $e$ are the electron mass and charge.} in  \cite{Borisov:1985ha,Erdas:2002wk,Bhattacharya:2008px} and for supercritical fields in \cite{Kuznetsov:2010sn}. The reaction (\ref{reaction}) reduces the neutrino mean free path to the values shorter than than the astrophysically relevant distances just after it leaves the regime of exponential suppression \cite{Erdas:2002wk}. This will produce a cutoff in the spectrum of neutrino sources if the latter possess strong magnetic fields in the region of neutrino emission.

We study the process (\ref{reaction}) using the  semiclassical ``worldline instanton'' method. This method is technically much simpler than the standard approach  \cite{Borisov:1985ha,Erdas:2002wk,Bhattacharya:2008px,Kuznetsov:2010sn}, based on the exact expressions for electron and W-boson wave functions (or propagators) in the magnetic field, and provides an independent verification of the results existing in the literature. 

%%%%%%%%%%%%%%%%%%

Worldline path integral approach \cite{feynman} is a powerful tool to study non-perturbative phenomena in quantum field theory, such as particle production in a classical external field.  The well-known example is the Schwinger effect \cite{schwinger} --- creation of electron-positron pairs from vacuum in a constant electric field. Affleck et. al. showed \cite{Affleck:1981bma} that the rate of the process can be expressed as the quantum mechanical partition function of an auxiliary system describing periodic motion of a charged particle in the external field, analytically continued to Euclidean time domain. The corresponding path integral  can be evaluated in the saddle point approximation. The method was generalized to pair production in time-dependent and space-dependent electric fields \cite{Dunne:2005sx}, including the case of pair production induced by a photon in the initial state \cite{Monin:2010qj}. In \cite{Satunin:2013an} this approach was used to study decay of photon to $e^+e^-$ pair in the magnetic field. It was also applied to reactions in theories beyond the Standard model such as monopole decay in electric field and W-boson decay to a monopole an dyon \cite{Monin:2005wz}, and to study particle production in de Sitter spacetime \cite{Guts:2013dha}. 

It was shown \cite{Affleck:1981bma,Dunne:2005sx}, that the exponential part of the rate of such Schwinger-like processes does not depend on the spin of charged particles (spin dependence appears only in the pre-exponential factor). So, for simplicity we will consider in our work all particles participating in the process as scalars.

\section{The width of neutrino decay $\nu\,\to\,e^-\,W^+$}

In this section we consider neutrino decay to an electron and $W^+$-boson in the external magnetic field. We are interested only in the main exponential behaviour of the result, which should be independent of particle spins.  
Instead of the electroweak theory for simplicity we consider a toy model with scalar particles. The Lagrangian of the model is
\begin{align}\label{Model}
\mathcal{L} = &D_\mu\phi^*D_\mu\phi - m_e^2\phi^*\phi + \notag\\
            + &D_\mu\chi^*D_\mu\chi - m_W^2\chi^*\chi + \\
            + &\frac{1}{2}\l \d_\mu\xi\r^2 -\frac{1}{4}F_{\mu\nu}F^{\mu\nu} + g\xi\phi^{*}\chi + h.c. \notag
\end{align}
Here $\xi$ is a real massless scalar field, representing ``neutrino'', $\phi$ and $\chi$ are scalar ``electron'' and ``W-boson''; $m_e$ and $m_W$ denote their masses. Interaction term includes constant $g$ of a mass dimension. Fields $\phi$ and $\chi$ interact with a gauge field $A_\mu$, which has the standard kinetic term. Covariant derivative $D_\mu$ is defined as usual, $D_\mu\phi=\l\d_\mu-ieA_\mu\r\phi$.  

In terms of our toy model we consider the process of a $\xi$ particle (neutrino) decay to a pair of $\phi$ particle and $\chi$ antiparticle (scalar electron and W-boson, respectively). Neutrino with four-momentum $k_\mu=(\omega,{\bf k})$  propagates orthogonally to the uniform magnetic field ${\bf H}$. We choose the coordinate system where the magnetic field is directed along the $x$-axis, neutrino momentum --- along the $y$-axis, so ${\bf k}=(0,\omega,0)$. The reaction is kinematically allowed  if $\omega > m_W + m_e$, we study it well above the threshold, $\omega \gg m_W + m_e$. Let us mention that all subsequent formulas are valid if the electron is replaced by muon or tau-lepton. 
%%%%%%%%

Following the optical theorem, the width of $\xi$ can be obtained from the imaginary part of its self-energy:
\be\label{Optical}
\Gamma = \frac{1}{2\omega}\mathrm{Im}\Sigma(k),
\ee
where $\Sigma(k)$ is the Fourier transform of the correlator:
\be\label{Pi1}
\Sigma(y-z)= \langle \chi^*(y)\phi(y)\phi^*(z)\chi(z)\rangle \; + \; \langle \phi^*(y)\chi(y)\chi^*(z)\phi(z)\rangle.
\ee
The first term in eq. (\ref{Pi1}) corresponds to creation of a $\phi$ particle and $\chi$ antiparticle; the second term --- to creation of a $\chi$ particle and $\phi$ antiparticle. We will concentrate only on the first term for two reasons. First, in the model (\ref{Model}) the exponential parts of both terms are equal, so for simplicity we can restrict to one of them. Second, in the realistic case of the Standard Model neutrino self-energy does not contain an analogy of the second term due to the lepton charge conservation.

The two-point Green function can be represented as (see \cite{Schubert:2001he}):
\begin{align} \label{correl}
& \langle \chi^*(y)\chi(z)\rangle = \int_0^\infty dT \langle y | \e^{-T\l D_\mu^2 + m_W^2\r} | z \rangle = \\
& = \int_0^\infty\frac{dT}{(2\pi T)^2}\e^{-m^2_W T}\frac{1}{N}\int_{x_\mu(0)=z_\mu}^{x_\mu(T)=y_\mu} Dx \e^{-\int_0^T\l \frac{\dot{x}_\mu^2}{4} - ieA_\mu\dot{x}_\mu\r d\tau}, \notag
\end{align}
where $N=\int_{x_\mu(0)=z_\mu}^{x_\mu(T)=y_\mu} Dx \e^{-\int_0^T\frac{\dot{x}_\mu^2}{4} d\tau}$ is a normalization factor. Summation over repeated Greek indices with Euclidean signature is understood. The notation $\dot{x}_\mu$ denotes derivative over $\tau$.
The formula similar to (\ref{correl}) is valid for the $\langle \phi^*(z)\phi(y)\rangle$ correlator.
Substituting both correlators into the self-energy (\ref{Pi1}), we obtain
\begin{align}\label{self-energy}
\Sigma(y-z) \; \propto \; &\int_0^\infty\frac{dT_1}{T_1^2}\int_0^\infty\frac{dT_2}{T_2^2}\e^{-m_e^2T_1-m_W^2T_2}\frac{1}{N_1N_2}\\
&\int_{p.b.c}Dx_\mu\delta^{(4)}\l x_\mu(0)-z_{\mu}\r\delta^{(4)}\l x_\mu(T_1)-y_{\mu}\r\e^{-\int_0^{T_1+T_2}\l \frac{\dot{x}_\mu^2}{4} - ieA_\mu\dot{x}_\mu\r d\tau}.  \notag
\end{align}
The notation 'p.b.c' means periodical boundary conditions $x_\mu(\tau)=x_\mu(\tau+T_1+T_2)$. $N_1$ and $N_2$ are the normalization factors connected with the two parts of the path integral.
Setting  $y_\mu+z_\mu=0$ and making the Fourier transformation of (\ref{self-energy}), we obtain
\be
\label{Gamma}
\Gamma \propto \frac{1}{\omega} \mathrm{Im} \int_0^\infty \frac{dT_1}{T_1^2}\int_0^\infty \frac{dT_2}{T_2^2} \frac{1}{N_1N_2}\int_{p.b.c} Dx_\mu \delta^{(4)}\l x_\mu(0)+x_{\mu}\l1/2\r\r\e^{-S[x_\mu]},
\ee
with
\begin{align}\label{Action1}
S \equiv S[x_\mu] & = m_e^2T_1 + \int_0^{1/2}\frac{\dot{x}_\mu^2}{8T_1}d\tau - ie\int_0^{1/2}A_\mu\dot{x}_\mu d\tau + \\
& +  m_W^2T_2 + \int_{1/2}^1\frac{\dot{x}_\mu^2}{8T_2}d\tau - ie\int_{1/2}^1A_\mu\dot{x}_\mu d\tau - ik_\mu^E\l x_\mu(1/2)-x_\mu(0)\r.
\notag
\end{align}
In the last expression we have rescaled the proper time $\tau$ so that the paths describing electron and W-boson correspond to $\tau \in \left[0,1/2\right]$ and $\tau \in \left[1/2,1\right]$ respectively.
The expression (\ref{Action1}) has the form of a sum of two Euclidean actions of two relativistic particles with different masses in the external electromagnetic field. Two sources of opposite signs located at the proper times $0$ and $1/2$, are added into the action. The strength of the sources depends on the neutrino momentum. We introduce the notation $k_\mu^E = \l i\omega,\vec{k}\r$. 

We expect that the integrals on the r.h.s. of (\ref{Gamma}) can be evaluated in the saddle point approximation. The saddle point equations for $x_\mu(\tau)$ give the classical trajectories (in general complex), which should be substituted into the action (\ref{Action1}). If the action (\ref{Action1}) on the solution is parametrically large, the width of the process is suppressed by the exponent of the action (with the minus sign). For further calculations it is convenient to choose the gauge $A_\mu=-\frac{1}{2}F_{\mu\nu}x_\nu$. Varying the action (\ref{Action1}) over $x_\mu$, we obtain the equations for different regions of parameter $\tau$ (we denote the solutions in the two regions $x_\mu^{(1)}$ and $x_\mu^{(2)}$);
\begin{align}
& \frac{\ddot{x}_\mu^{(1)}}{4T_1}=ieF_{\mu\nu}\dot{x}_\nu^{(1)}, \qquad 0<\tau<1/2,  \label{DirectEq} \\
& \frac{\ddot{x}_\mu^{(2)}}{4T_2}=ieF_{\mu\nu}\dot{x}_\nu^{(2)}, \qquad 1/2<\tau<1, \label{ReversedEq} 
\end{align}
and boundary conditions at the points $0$ and $1/2$:
\begin{align}
& \frac{\dot{x}_\mu^{(1)}(1/2)}{T_1} - \frac{\dot{x}_\mu^{(2)}(1/2)}{T_2} = 4ik_\mu^E, \label{bc1}\\
& \frac{\dot{x}_\mu^{(2)}(0)}{T_2} - \frac{\dot{x}_\mu^{(1)}(0)}{T_1} = - 4ik_\mu^E. \label{bc2}
\end{align}

We are looking for a solution of equations (\ref{DirectEq})---(\ref{bc2}) that describes a closed trajectory in four-dimensional complex spacetime --- ``worldline instanton''. The solution is composed of two different hyperbolic arcs, defined on the segments $\tau \in (0,1/2)$ and $\tau \in (1/2,1)$, respectively:
\begin{align} \label{s1}
& x_0^{(1)} = A_0 \l \tau - \frac{1}{4}\r, \notag \\
& x_2^{(1)} = iA_L \sh\l 4\theta_1\l \tau - \frac{1}{4}\r\r,  \\
& x_3^{(1)} = A_L \left[ \ch\l 4\theta_1\l \tau - \frac{1}{4}\r\r - \ch\theta_1\right], \notag
\end{align}
and
\begin{align} \label {s2}
& x_0^{(2)} = -A_0 \l \tau - \frac{3}{4}\r, \notag \\
& x_2^{(2)} = -iA_R \sh\l 4\theta_2\l \tau - \frac{3}{4}\r\r,  \\
& x_3^{(2)} = -A_R \left[ \ch\l 4\theta_2\l \tau - \frac{3}{4}\r\r - \ch\theta_2\right].\notag
\end{align}
Here, for simplicity, instead of $T_i$ we use dimensionless parameters $\theta_i=T_ieH$. Other parameters are determined in the following way:
$$
A_0 = \frac{4\omega}{eH}\cdot \frac{\theta_1\theta_2}{\theta_1+\theta_2}, \qquad A_L = \frac{\omega}{eH}\cdot \frac{\sh\theta_2}{\sh(\theta_1+\theta_2)}, \qquad A_R = \frac{\omega}{eH}\cdot \frac{\sh\theta_1}{\sh(\theta_1+\theta_2)}.
$$
This solution is shown in Fig.1.
\begin{figure}[h]
\begin{center}
\begin{tabular}{cc}
%\begin{minipage}[h]{0.48\linewidth}
\includegraphics[width=0.48\linewidth]{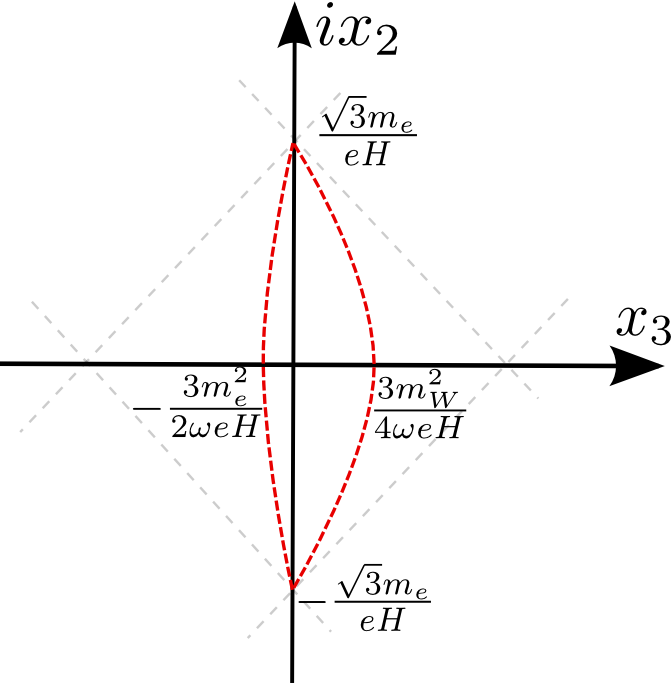}
%\caption{"Worldline instanton" --- the classical trajectory describing neutrino decay ($\theta_1 \ll 1$). The projection of the trajectory on the plane $(ix_2,x_3)$ is shown.} 
%\label{winstA} 
%\end{minipage}
%\hfill
%\begin{minipage}[h]{0.48\linewidth}
&
\includegraphics[width=0.48\linewidth]{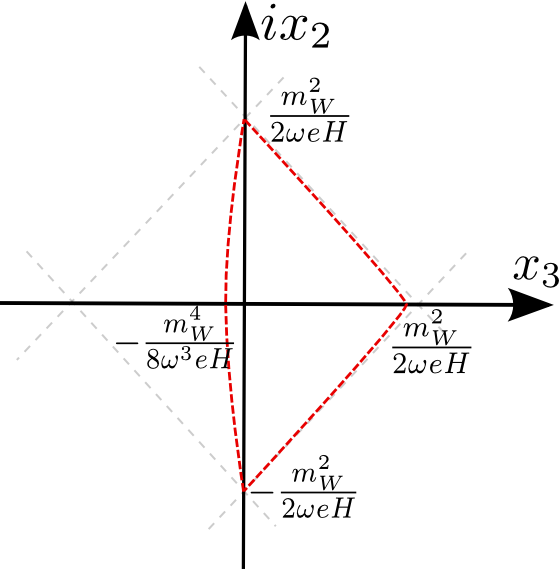}
%\caption{"Worldline instanton" --- the classical trajectory describing neutrino decay  ($\theta_1 \gg 1$). The projection of the trajectory on the plane $(ix_2,x_3)$ is shown.} 
%\label{winstB}
%\end{minipage}
\end{tabular}
\caption{"Worldline instanton": the classical trajectory describing neutrino decay. The left (right) panel refers to the case $\theta_1 \ll 1$  ($\theta_1 \gg 1$). The projection of the trajectory on the plane $(ix_2,x_3)$ is shown.} 
\end{center}
\end{figure}
Substituting solution (\ref{s1}),(\ref{s2}) into (\ref{Action1}), we obtain the action on the classical solution:

\be\label{ActionT}
S\left[ \theta_1,\theta_2\right] = \frac{m_e^2}{eH}\theta_1 + \frac{m_W^2}{eH}\theta_2 + \frac{\omega^2}{eH}\left[  -  \frac{\theta_1\theta_2}{\theta_1+\theta_2} + \frac{\sh\theta_1\sh\theta_2}{\sh(\theta_1+\theta_2)}\right]. 
\ee

Having performed the path integral we are left with two ordinary integrals over  $\theta_1$ and $\theta_2$:
$\int_0^\infty d\theta_1\int_0^\infty d\theta_2\theta_1^{-2}\theta_2^{-2}\e^{-S\left[ \theta_1,\theta_2\right]}.$
As the classical solution is assumed to be large (which will be checked a posteriori), these integrals can also be evaluated with the saddle point method. 
Varying the action (\ref{ActionT}) over $\theta_1,\ \theta_2$ we obtain:
\begin{align}
\frac{\sh^2\theta_2}{\sh^2(\theta_1+\theta_2)} = \frac{\theta_2^2}{(\theta_1+\theta_2)^2} - \frac{m_e^2}{\omega^2},\label{e1}\\
\frac{\sh^2\theta_1}{\sh^2(\theta_1+\theta_2)} = \frac{\theta_1^2}{(\theta_1+\theta_2)^2} - \frac{m_W^2}{\omega^2}.\label{e2}
\end{align}
If the masses of final particles are equal, $m_W=m_e$, these equations reduce to those studied in \cite{Satunin:2013an} in the context of photon decay into electron-positron pair; in this case $\theta_1=\theta_2$. For general values of $m_W$, $m_e$ it is impossible to solve eqs.(\ref{e1}),(\ref{e2}) at arbitrary values of $\omega$ and $H$ analytically. However, if $m_W \gg m_e$ approximate analytical solution exists in two regimes. First, suppose $\theta_1$ and $\theta_2$ are small compared to unity and assume the hierarchy $\theta_2 \ll \theta_1 \ll 1$. Then from eqs. (\ref{e1}),(\ref{e2}) we obtain the saddle point values of $\theta_1$ and $\theta_2$:
\be\label{STA}
\theta_1^c=\frac{\sqrt{3}}{2}\frac{m_W^2}{m_e\omega}, \qquad \theta_2^c=\sqrt{3}\frac{m_e}{\omega}.
\ee
The configuration of the instanton in this case is shown on the left panel of Fig.1. The auxiliary time $\tau$ increases anticlockwise. The instanton consists of two smooth hyperbolic arcs with the left arc being the trajectory of the W-boson and the right arc --- of the electron. Being continued to Minkowski spacetime, the closed trajectory transforms to real trajectories of outgoing particles. In this picture electron, being ultrarelativistic, carries most of the neutrino energy, in agreement with the standard consideration \cite{Kuznetsov:2013sea}.

In order to obtain the neutrino width we substitute the solution (\ref{STA}) to the action\footnote{Interestingly, this action is equal to the area enclosed by the classical trajectory on the $(ix_2,x_3)$ plane multiplied by $eH$.} (\ref{ActionT}):
\be \label{SA}
S=\frac{\sqrt{3}m_em_W^2}{\omega eH}.
\ee
If the action (\ref{SA}) is parametrically large, $S \gg 1$, the neutrino width is proportional to the minus exponent of the action:
\be\label{A}
\Gamma \propto \e^{-\frac{\sqrt{3}m_em_W^2}{\omega eH}}.
\ee

The pre-exponential factor can be obtained from the Gaussian integration over the small fluctuations around the worldline instanton. Such calculation is beyond the scope of the present work. We limit ourselves to the proof in Appendix A that these fluctuations contain a single negative mode which renders the contribution of the worldline instanton into the self-energy $\Sigma(k)$ purely imaginary \cite{Callan:1977pt} and hence its contribution into the decay width (\ref{Optical}) is indeed nonzero.

The formula (\ref{A}) is valid within the following approximations: the condition $\theta_1^c \ll 1$ gives $\omega \gg \frac{\sqrt{3}m_W^2}{2m_e}$ while the semiclassical limit $S\gg 1$ requires $\omega \ll \frac{\sqrt{3}m_em_W^2}{eH}$. These two conditions are simultaneously fulfilled in a certain region on the $(\omega,H)$ plane (see Fig. 2).  Inside this region eq. (\ref{A}) coincides with the results of the previous studies \cite{Borisov:1985ha, Erdas:2002wk} (see also \cite{Kuznetsov:2013sea}). Note that this region lies entirely in subcritical magnetic fields, $H \ll m_e^2/e$.  From the physical viewpoint the condition $S \sim 1$, or
\be\label{line1}
\omega H \sim \sqrt{3}m_em_W^3/e
\ee
determines the effective threshold of the reaction (\ref{reaction}) in subcritical magnetic field.

To address the case of supercritical magnetic fields, that are believed to exist in magnetars, we must look for other solutions of eqs. (\ref{e1}), (\ref{e2}). From (\ref{line1}) we observe that for larger magnetic fields the regime of exponential suppression shifts to lower neutrino energies. On the other hand, decreasing neutrino energy in (\ref{STA}), we can violate the condition $\theta_1 \ll 1$ still being in the semiclassical regime. Hence we are led to consider the case of large $\theta_1$: 
$$
\theta_2 \ll 1 \ll \theta_1.
$$
Solving equations (\ref{e1}),(\ref{e2}) in this limit, we obtain
\be\label{Reg2}
\theta_1^c=\frac{m_W^2}{2m_e\omega}, \qquad \theta_2^c=\frac{m_W^2}{2\omega^2}.
\ee
Note that though the formula for $\theta_1$, up to a numerical factor, remains the same as in (\ref{STA}), the expression for $\theta_2$ changes. 
Substituting the solution (\ref{Reg2}) into the action (\ref{ActionT}), we obtain\footnote{Again, the action is equal to the area enclose by worldline instanton times $eH$.}
\be\label{SB}
S=\frac{m_W^4}{4\omega^2 eH}.
\ee
Unlike the previous case, the right arc of the trajectory is highly curved and is very close to the lightcone. In fact, the parameters of the trajectory in the leading order (see Fig. 1, right panel) and the exponent (\ref{SB}) do not depend on the electron mass. Thus, the case $\theta_1 \gg 1$ corresponds to the limit of massless electron.

\begin{figure}[h]
\begin{center}
\includegraphics[width=0.99\linewidth]{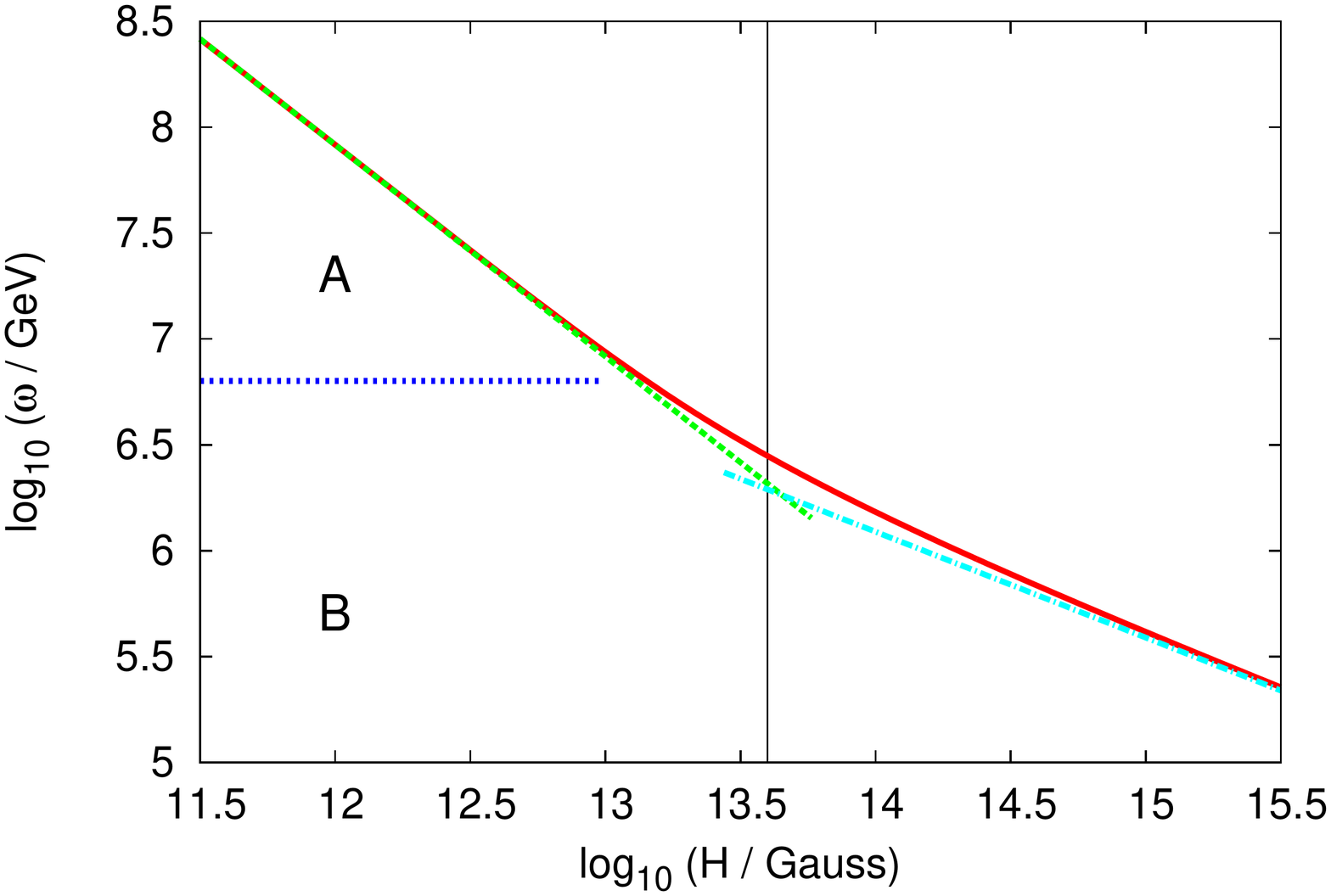}
\caption{Regimes of the reaction $\nu\to e^-W^+$ at different values of neutrino energy $\omega$ and magnetic field $H$. The solid line denotes the effective threshold of the reaction: in the region below if the reaction is exponentially suppressed while above this line the suppression disappears. Dashed and dashed-dotted lines show asymptotic, described by eqs (\ref{line1}) and (\ref{line2}) respectively. Vertical line at $H \sim 10^{13.6}$ G denotes critical value of the magnetic field. The reaction width is determined by formula (\ref{A}) (by formula (\ref{GB})) in the region A (B).}
\end{center}
\end{figure}

In the semiclassical regime $S \gg 1$ the decay width becomes
\be\label{GB}
\Gamma \propto \e^{-\frac{m_W^4}{4\omega^2 eH}}.
\ee
This result is obtained under two approximations: applicability of the semiclassical expansion requires $\omega \ll \frac{m_W^2}{2\sqrt{eH}}$ while the condition $\theta_1^c \gg 1$ gives $\omega \ll \frac{m_W^2}{2m_e}$. These conditions are satisfied in a wide region (see Fig.2),  containing both sub- and supercritical magnetic fields. The formula (\ref{GB}) was previously obtained Ref. \cite{Kuznetsov:2010sn} by direct calculations in the limit of massless electron. The calculation relies on the approximation $H \gg m_e^2/e$. We have found that the formula (\ref{GB}) is valid in a larger region of parameters that what was argued in  \cite{Kuznetsov:2010sn}. Note, however, that the additional region $eH \ll m_e^2$, $\omega \ll m_W^2/2m_e$ corresponds to large exponential suppression, so it presents only academic interest. From the practical viewpoint, the effective threshold of the neutrino decay in supercritical fields is determined by 
\be\label{line2}
\omega \sim \frac{m_W^2}{2\sqrt{eH}},
\ee
in agreement with \cite{Kuznetsov:2010sn}. 

In order to compute the effective threshold of neutrino decay for magnetic field compatible to the critical value we solve the system (\ref{e1}), (\ref{e2}) numerically (see the solid line at the Fig. 2). The behaviour of the effective threshold agrees with the results of Kuznetsov et.al. \cite{Kuznetsov:2010rg}.

\section{Discussion}

We have shown that the worldline instanton method can be applied to the calculation of the neutrino decay rate into electron and W-boson in the external magnetic field in the regime when this rate is exponentially suppressed. We have obtained analytic expressions for the suppression exponent in two limiting cases; these expressions smoothly match along the boundary $\omega \sim m_W^2/m_e$ on the $(\omega,H)$ plane.

Our approach provides a technically simple derivation of the effective threshold energy for the reaction where the exponential suppression disappears, as a function of the magnetic field.
These estimates must be taken into account in the analysis of the models for astrophysical sources of very high energy neutrinos. For example, from Fig. 2 we see that neutrinos with energies higher than $10^{15}$ eV cannot escape from the vicinity of a magnetar with the magnetic field exceeding a few times $10^{14}$ G.

It is straightforward to generalize  our approach to a wide class of processes. In appendix B we use it to estimate the effective threshold for the reaction $\nu \to \nu e^+e^-$. It can be also applied to the case of more complicated field configuration,  such as crossed electric and magnetic fields. The latter configuration can be realized in fast spinning magnetized astrophysical objects like pulsars or black holes at transient periods then the electric field is not screened by the surrounding plasma. 

It is worth stressing that in our calculation we never made use of the precise properties of neutrino, electron and W-boson --- in fact, we substituted them for simplicity by scalar particles with cubic interaction. Therefore, it will apply almost without changes to a decay of a neutral particle into two charged ones in the magnetic field in theories beyond the Standard Model. Examples where this process can be relevant include models with axions \cite{Mikheev:1997cr}, paraphotons \cite{Ahlers:2007rd}, particle decays to millicharged particles \cite{Davidson:2000hf} e.t.c.  

Furthermore,  it was shown \cite{Satunin:2013an} that the method of "worldline instantons" can be easily generalized to theories with violation of Lorentz invariance. The exponential suppression of the decay is sensitive to kinematics, so minor deviation of the particle dispersion relations from the relativistic form can change the decay width significantly.

As discussed in \cite{Rubtsov:2013wwa}, this will lead to very strong constraints on such observation for electrons, positrons and photons if the decay of ultra-high-energy photons in the geomagnetic field is observed in future.
It is straightforward to incorporate effects of Lorentz invariance violation in the calculation of the present paper along the lines of  \cite{Satunin:2013an}. However, at the moment one does not expect to obtain any useful constraints on these effects from neutrino astrophysics due to large uncertainties in the neutrino source model.
 
\paragraph*{Acknowledgements}

The author thanks Alexander Kuznetsov, Grigory Rubtsov, Slava Rychkov and especially Sergey Sibiryakov for helpful discussions. This work was supported by RSF grant 14-12-01340. The author thanks CERN TH group for hospitality. 

\appendix
\section{The negative mode}

To complete the calculation of neutrino decay we should check that the integrals in the expression (\ref{Gamma}) have nonzero imaginary part. For this reason we show that fluctuations near the saddle classical solution contain a single negative mode along which the action (\ref{Action1}) decreases. Thus, according to the standard arguments \cite{Callan:1977pt}, this lead to the appearance of a factor $i/2$ in front of the path integral. 

First, consider  small fluctuations $\delta x_\mu$ near the classical solution (\ref{s1})-(\ref{s2}) in the action (\ref{Action1}). The action is quadratic in $x_\mu$, so its second variation does not depend on the classical solution. Fluctuations $\delta x_\mu(\tau)$ at $\tau\neq 0,1/2$ contribute an exactly positive Gaussian integral. The second variation of the action along the remaining the mode $\xi_\mu = \sqrt{eH}\l \delta x_\mu(0) - \delta x_\mu(1/2)\r$ can be written as 
$$
\delta^2 S=\left[ \frac{\xi_0^2+\xi_1^2}{\theta_1} + \frac{\xi_2^2+\xi_3^2}{\sh^2\theta_1} \l 2\theta_1 - \sh\theta_1\ch\theta_1\r \right] + \left[ \frac{\xi_0^2+\xi_1^2}{\theta_2} + \frac{\xi_2^2+\xi_3^2}{\sh^2\theta_2} \l 2\theta_2 - \sh\theta_2\ch\theta_2\r \right].
$$
For the saddle-point values we always have $\theta_2 \ll 1$, $\theta_2 \ll \theta_1$. Hence, the fluctuations of the action along the four modes $\xi_\mu$ are positive. There are no zero modes due to the delta-function in (\ref{Gamma}). 

Next, turn to the fluctuations of $\theta_i$ and consider the cases of small and large $\theta_1$ separately.
In the limit $\theta_1 \ll 1$ the second variation of the action (\ref{ActionT}) becomes
$$
\delta^2S=-\frac{2\omega^2}{3eH}\theta_1^c\l \delta\theta_2 +\delta\theta_1\frac{\theta^c_2}{\theta^c_1} \r^2 +  \frac{2\omega^2}{3eH}\frac{\l\theta_2^c\r^2}{\theta_1^c}\l\delta\theta_1\r^2 .
$$
Thus, $\l \delta\theta_2 + \delta\theta_1\cdot\theta_2^c/\theta_1^c\r$ is a negative mode. Note that this mode is mostly associated with the fluctuation $\delta\theta_2$ because of the relation $\theta_2^c/\theta_1^c$ is small.
Consider the opposite case $\theta_1^c \gg 1$. The second variation of the action diagonalizes in the following way:
$$
\delta^2 S = -\frac{2\omega^2}{eH}\l \delta\theta_2\r^2 + \frac{2\omega^2}{\theta_1^ceH}\l \frac{\theta_2^c}{\theta_1^c}\delta\theta_1 - \delta\theta_2\r^2.
$$
Hence, in this case we again have a negative mode $\delta\theta_2$. 

It is worth to point out that $\theta_2$ sets the linear extent of the instanton along the $ix_2$ direction in both cases (see eqs (\ref{s1}),(\ref{s2})). Thus we conclude with negative mode correspondence to the change of the overall instanton size, similar to other studies using the worldline instanton approach.

\section{$e^+e^-$ pair production by neutrinos}

The worldline instanton approach to particle decay reactions can be applied as well to particle decays with three particles in the final state, if one of these particles is neutral. 
In particular, the rate of the process $\nu \to \nu e^+e^-$ in the magnetic field (the exponential part) can be easily calculated using the known rate of photon decay in magnetic field.

Consider neutrino with energy $\omega$, producing electron-positron pair in magnetic field ${\bf H}$ (as usual, for simplicity we consider neutrino momentum orthogonal to the field); the remaining neutrino carries energy $\omega'$. 
The situation is the same as if a single massless particle with energy $(\omega-\omega')$ decayed into electron-positron. The rate of the latter process was previously obtained by worldline instanton method \cite{Satunin:2013an}. Integrating over $\omega'$, we obtain
$$
\Gamma \propto \int_0^{\omega-2m_e} d\omega'\; \e^{-\frac{8m_e^3}{3(\omega-\omega')eH}} \;\;\propto\;\; \e^{-\frac{8m_e^3}{3\omega eH}}.
$$
Here we take care only of the leading exponential factor. This simple estimate is in agreement with the previous studies \cite{Borisov:1993rba, Kuznetsov:2000gn}. The transition of this reaction from exponentially suppressed regime to non-suppressed one can again be considered as an effective threshold. However, at least in the case of subcritical magnetic field this reaction is relatively weak \cite{Kuznetsov:1996vy} even in the absence of the exponential suppression: the neutrino mean free path is greater than typical length of the strong magnetic field in astrophysical objects (see Fig.2 in \cite{Erdas:2002wk}).


\begin{thebibliography}{99}

%\cite{Risse:2007sd}
\bibitem{Risse:2007sd}
  M.~Risse and P.~Homola,
  %``Search for ultrahigh energy photons using air showers,''
  Mod.\ Phys.\ Lett.\ A {\bf 22} (2007) 749
  [astro-ph/0702632 [ASTRO-PH]].
  %%CITATION = ASTRO-PH/0702632;%%

%\cite{Aartsen:2013bka}
\bibitem{Aartsen:2013bka}
  M.~G.~Aartsen {\it et al.}  [IceCube Collaboration],
  %``First observation of PeV-energy neutrinos with IceCube,''
  Phys.\ Rev.\ Lett.\  {\bf 111} (2013) 021103
  [arXiv:1304.5356 [astro-ph.HE]].
  %%CITATION = ARXIV:1304.5356;%%
  %26 citations counted in INSPIRE as of 15 Aug 2013

%\cite{Erdas:1990gy}
\bibitem{Erdas:1990gy}
  A.~Erdas and G.~Feldman,
  %``Magnetic Field Effects On Lagrangians And Neutrino Selfenergies In The Salam-weinberg Theory In Arbitrary Gauges,''
  Nucl.\ Phys.\ B {\bf 343} (1990) 597.
  %%CITATION = NUPHA,B343,597;%%
  %31 citations counted in INSPIRE as of 28 Aug 2013

%\cite{Borisov:1985ha}
\bibitem{Borisov:1985ha}
  A.~V.~Borisov, V.~C.~Zhukovsky, A.~V.~Kurilin and A.~I.~Ternov,
  %``Radiative Corrections To Neutrino Mass In External Electromagnetic Field. (in Russian),''
  Yad.\ Fiz.\  {\bf 41} (1985) 743.
  %%CITATION = YAFIA,41,743;%%
  %18 citations counted in INSPIRE as of 17 Jul 2013

%\cite{Erdas:2002wk}
\bibitem{Erdas:2002wk}
  A.~Erdas and M.~Lissia,
  %``High-energy neutrino conversion into electron W pair in magnetic field and its contribution to neutrino absorption,''
  Phys.\ Rev.\ D {\bf 67} (2003) 033001
  [hep-ph/0208111].
  %%CITATION = HEP-PH/0208111;%%
  %4 citations counted in INSPIRE as of 07 Aug 2013

%\cite{Bhattacharya:2008px}
\bibitem{Bhattacharya:2008px}
  K.~Bhattacharya and S.~Sahu,
  %``Neutrino absorption by W production in the presence of a magnetic field,''
  Eur.\ Phys.\ J.\ C {\bf 62} (2009) 481
  [arXiv:0811.1692 [hep-ph]].
  %%CITATION = ARXIV:0811.1692;%%
  %2 citations counted in INSPIRE as of 07 Aug 2013


%\cite{Kuznetsov:2010sn}
\bibitem{Kuznetsov:2010sn}
  A.~V.~Kuznetsov, N.~V.~Mikheev and A.~V.~Serghienko,
  %``High energy neutrino absorption by W production in a strong magnetic field,''
  Phys.\ Lett.\ B {\bf 690} (2010) 386
  [arXiv:1002.3804 [hep-ph]].
  %%CITATION = ARXIV:1002.3804;%%
  %1 citations counted in INSPIRE as of 17 Jul 2013


%\cite{Borisov:1993rba}
\bibitem{Borisov:1993rba}
  A.~V.~Borisov, V.~C.~Zhukovsky and A.~I.~Ternov,
  %``Electron positron pair production by a neutrino in an external electromagnetic field,''
  Phys.\ Lett.\ B {\bf 318} (1993) 489.
  %%CITATION = PHLTA,B318,489;%%
  %16 citations counted in INSPIRE as of 15 Jul 2014

%\cite{Kuznetsov:1996vy}
\bibitem{Kuznetsov:1996vy}
  A.~V.~Kuznetsov and N.~V.~Mikheev,
  %``The Neutrino energy and momentum loss through the process neutrino ---> neutrino e- e+ in a strong magnetic field,''
  Phys.\ Lett.\ B {\bf 394} (1997) 123
  [hep-ph/9612312].
  %%CITATION = HEP-PH/9612312;%%
  %28 citations counted in INSPIRE as of 17 Jul 2014

%\cite{Kuznetsov:2000gn}
\bibitem{Kuznetsov:2000gn}
  A.~V.~Kuznetsov, N.~V.~Mikheev and D.~A.~Rumyantsev,
  %``Lepton pair production by high-energy neutrino in an external electromagnetic field,''
  Mod.\ Phys.\ Lett.\ A {\bf 15} (2000) 573
  [hep-ph/0003216].
  %%CITATION = HEP-PH/0003216;%%
  %2 citations counted in INSPIRE as of 15 Jul 2014


%\cite{Kuznetsov:2013sea}
\bibitem{Kuznetsov:2013sea}
  A.~Kuznetsov and N.~Mikheev,
  %``Electroweak Processes in External Active Media,''
  Springer Tracts Mod.\ Phys.\  {\bf 252} (2013).
  %%CITATION = STPHB,252,pp 1;%%


%\cite{feynman}
\bibitem{feynman}
R.P. Feynman, 
``Mathematical formulation of the quantum theory of electromagnetic interaction'',
Phys. Rev. {\bf 80} (1950) 440;
``An Operator Calculus Having Applications in Quantum Electrodynamics'',
Phys. Rev. {\bf 84} (1951) 108.

%\cite{schwinger}
\bibitem{schwinger}
J. Schwinger, 
``On gauge invariance and vacuum polarization'', 
Phys. Rev.
{\bf 82}, 664 (1951).

%\cite{Affleck:1981bma}
\bibitem{Affleck:1981bma}
  I.~K.~Affleck, O.~Alvarez and N.~S.~Manton,
  %``Pair Production At Strong Coupling In Weak External Fields,''
  Nucl.\ Phys.\ B {\bf 197} (1982) 509.
  %%CITATION = NUPHA,B197,509;%%

%\cite{Dunne:2005sx}
\bibitem{Dunne:2005sx}
  G.~V.~Dunne and C.~Schubert,
  %``Worldline instantons and pair production in inhomogeneous fields,''
  Phys.\ Rev.\ D {\bf 72} (2005) 105004
  [hep-th/0507174].
  %%CITATION = HEP-TH/0507174;%%

%\cite{Monin:2010qj}
\bibitem{Monin:2010qj}
  A.~Monin and M.~B.~Voloshin,
  %``Semiclassical Calculation of Photon-Stimulated Schwinger Pair Creation,''
  Phys.\ Rev.\ D {\bf 81} (2010) 085014
  [arXiv:1001.3354 [hep-th]].
  %%CITATION = ARXIV:1001.3354;%%  

%\cite{Satunin:2013an}
\bibitem{Satunin:2013an}
  P.~Satunin,
  %``Width of photon decay in a magnetic field: Elementary semiclassical derivation and sensitivity to Lorentz violation,''
  Phys.\ Rev.\ D {\bf 87} (2013) 10,  105015
  [arXiv:1301.5707 [hep-th]].
  %%CITATION = ARXIV:1301.5707;%%
  %5 citations counted in INSPIRE as of 17 Jul 2014

%\cite{Monin:2005wz}
\bibitem{Monin:2005wz}
  A.~K.~Monin,
  %``Monopole decay in the external electric field,''
  JHEP {\bf 0510} (2005) 109
  [hep-th/0509047].
  %%CITATION = HEP-TH/0509047;%%
  %7 citations counted in INSPIRE as of 28 Aug 2013

%\cite{Guts:2013dha}
\bibitem{Guts:2013dha}
  S.~Guts,
  ``Semiclassical treatment of pair creation in de Sitter space,''
  arXiv:1312.2429 [hep-ph].
  %%CITATION = ARXIV:1312.2429;%%

%\cite{Schubert:2001he}
\bibitem{Schubert:2001he}
  C.~Schubert,
  %``Perturbative quantum field theory in the string inspired formalism,''
  Phys.\ Rept.\  {\bf 355} (2001) 73
  [hep-th/0101036].
  %%CITATION = HEP-TH/0101036;%%
  %169 citations counted in INSPIRE as of 28 Nov 2013



%\cite{Callan:1977pt}
\bibitem{Callan:1977pt}
  C.~G.~Callan, Jr. and S.~R.~Coleman,
  %``The Fate of the False Vacuum. 2. First Quantum Corrections,''
  Phys.\ Rev.\ D {\bf 16} (1977) 1762.
  %%CITATION = PHRVA,D16,1762;%%
  %857 citations counted in INSPIRE as of 12 Mar 2014


%\cite{Kuznetsov:2010rg}
\bibitem{Kuznetsov:2010rg}
  A.~V.~Kuznetsov, N.~V.~Mikheev and A.~V.~Serghienko,
  ``A Decay of the ultra-high-energy neutrino $\nu_e \to e^- W^+$ in a magnetic field and its influence on the shape of the neutrino spectrum,'' Proceedinds of the XVI International Seminar Quarks'2010, Kolomna,
  arXiv:1010.0582 [hep-ph].
  %%CITATION = ARXIV:1010.0582;%%

%\cite{Mikheev:1997cr}
\bibitem{Mikheev:1997cr}
  N.~V.~Mikheev and L.~A.~Vassilevskaya,
  %``About an influence of external electromagnetic fields on the axion decays,''
  Phys.\ Atom.\ Nucl.\  {\bf 61} (1998) 1041
   [Yad.\ Fiz.\  {\bf 61} (1998) 1135]
  [hep-ph/9708293].
  %%CITATION = HEP-PH/9708293;%%

%\cite{Ahlers:2007rd}
\bibitem{Ahlers:2007rd}
  M.~Ahlers, H.~Gies, J.~Jaeckel, J.~Redondo and A.~Ringwald,
  %``Light from the hidden sector,''
  Phys.\ Rev.\ D {\bf 76} (2007) 115005
  [arXiv:0706.2836 [hep-ph]].
  %%CITATION = ARXIV:0706.2836;%%
  %84 citations counted in INSPIRE as of 22 Jul 2014


%\cite{Davidson:2000hf}
\bibitem{Davidson:2000hf}
  S.~Davidson, S.~Hannestad and G.~Raffelt,
  %``Updated bounds on millicharged particles,''
  JHEP {\bf 0005} (2000) 003
  [hep-ph/0001179].
  %%CITATION = HEP-PH/0001179;%%
  %170 citations counted in INSPIRE as of 15 Jul 2014



%\cite{Rubtsov:2013wwa}
\bibitem{Rubtsov:2013wwa}
  G.~Rubtsov, P.~Satunin and S.~Sibiryakov,
  %``Prospective constraints on Lorentz violation from ultrahigh-energy photon detection,''
  Phys.\ Rev.\ D {\bf 89} (2014) 123011
  [arXiv:1312.4368 [astro-ph.HE]].
  %%CITATION = ARXIV:1312.4368;%%
  %1 citations counted in INSPIRE as of 24 Jul 2014





%%%%%%%%%%%%%%%%%%%%%%%%%%%%%%%%%%%%%%%%%%%%%%%%%%%%%%%%%%%%%%%%%%%%5

%\cite{Selivanov:1985vt}
%\bibitem{Selivanov:1985vt}
 % K.~B.~Selivanov and M.~B.~Voloshin,
  %``Destruction Of False Vacuum By Massive Particles,''
 % JETP Lett.\  {\bf 42} (1985) 422.
  %%CITATION = JTPLA,42,422;%%
  %24 citations counted in INSPIRE as of 28 Aug 2013








\end{thebibliography}
\end{document}